\documentclass[conference]{IEEEtran}
\usepackage{amsmath}
\usepackage[cmintegrals]{newtxmath}
\usepackage{url}
\usepackage{soul,color,graphicx,float,epstopdf}
\usepackage{tablefootnote,textcomp,gensymb}
\usepackage{eurosym,booktabs,multirow}
\usepackage{subcaption}
\usepackage{amsfonts} 
\usepackage{algorithmic}
\usepackage{array}
\usepackage{stfloats}
\usepackage{float}
\usepackage{mathtools}
\usepackage{graphicx}
\usepackage{pdfpages}
\usepackage{balance}
\usepackage{xcolor}
\usepackage{comment}
\usepackage[normalem]{ulem}
\usepackage{dsfont}
\usepackage[linesnumbered, ruled]{algorithm2e}
\usepackage{enumitem}
\usepackage{tikz}
\usepackage[implicit=false]{hyperref}
\usetikzlibrary{arrows,shapes,positioning,shadows,trees}

\usepackage{verbatim}
\usepackage{cite}

\tikzset{
  basic/.style  = {draw, text width=4cm, drop shadow, font=\sffamily, rectangle},
  root/.style   = {basic, rounded corners=2pt, thin, align=center,
                   fill=red!20},
  level 2/.style = {basic, rounded corners=6pt, thin,align=center, fill=blue!20,
                   text width=9em},
  level 3/.style = {basic, thin, align=left, fill=pink!20, text width=8em}
}

\definecolor{Orange}{rgb}{1,0.5,0}

\newcommand*{\rom}[1]{\expandafter\@slowromancap\romannumeral #1@}

\newcommand{\todo}[1]{\textsf{\textbf{\textcolor{Orange}{[#1]}}}}

\begin{document}
\title{Demo: Integration of Marketplace for \\the 5G Open RAN Ecosystem}
\author{\IEEEauthorblockN{
Tim Farnham,
Sajida Gufran,
Peizheng Li,
Adnan Aijaz
}\\ 
\vspace{-2.00mm}
\IEEEauthorblockA{
\IEEEauthorrefmark{0} Bristol Research and Innovation Laboratory, Toshiba Europe Ltd., U.K.\\
Email: {\{firstname.lastname\}@toshiba-bril.com}
}}

\maketitle
\begin{abstract}
The Open RAN API and interface standards facilitate the new ecosystems where distinct hardware and software components are brought together to build 5G systems. Key to this concept is the seamless and efficient integration and monetization process among stakeholders. A marketplace serves as a means to realize this collaborative revenue sharing, eliminating the need for intricate proprietary agreements or contracts between each participant. This demo presents the marketplace strategy emphasizing software integration across diverse deployment settings, utilizing the API-centric integration Platform-as-a-Service (iPaas) model aligned with Open RAN standards.
\end{abstract}

\section{Introduction}
The monetization of 5G Open RAN systems is pivotal for open eco-system success, enabling tailored deployment and intelligent control~\cite{10.3389/frcmn.2023.1127039}. Certain current marketplace models concentrate on specific use-cases and services. For instance, the NECOS\cite{H2020-NECOS} marketplace employs a Slice as a Service, encompassing resource requests and their alignment with resources from various tenants and providers. 
The marketplace of 5GTANGO \cite{5GTANGO} appears as the concept of a store with a customizable orchestrator, network slice manager and slice-to-network-service-mapper.
In contrast, our proposed marketplace approach follows an integration Platform-as-a-Service (iPaaS) paradigm that permits an open API-centric monetization. This is agnostic to the specific services, applications (x/rApps) and environments in which they are deployed, rather than support monetization based on different business models through fine-grained API access control and monitoring. This can be subscription based with pay-per-use or quota limits and also with optional service or performance level agreements and differentiation.  
This demo shows the API-centric iPaaS marketplace approach supporting Open RAN APIs, to deploy 5G digital twins (DT), applications (x/rApps) and services within an overall 5G Open RAN infrastructure with associated RAN Intelligent Controllers (RICs). 

\section{Design and implementation}
\subsection{Marketplace Approach}
\begin{figure}[t]   
        \centering   
        \includegraphics[width=0.8\columnwidth]{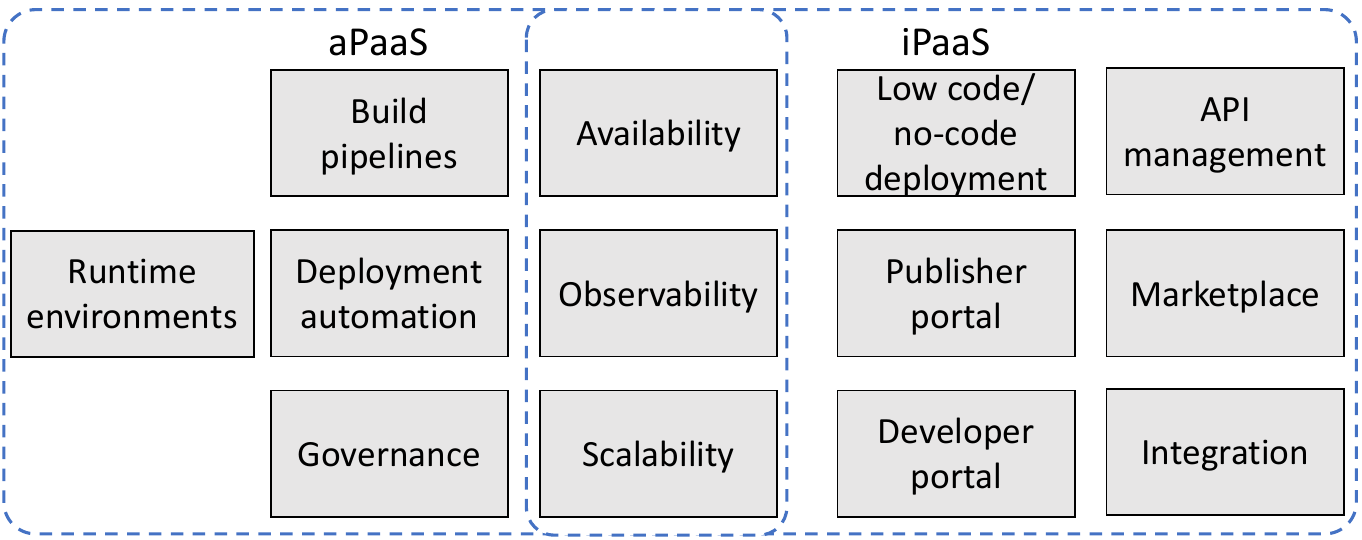}       \caption{Platform approaches comparing aPaaS and iPaaS.} \label{fig:platforms}
            \vspace{-5.00mm}
\end{figure}

There are diverse marketplace approaches available, such as iPaaS, application Platform-as-a-Service (aPaaS) and Software-as-a-Service (SaaS). Lately, telecommunication-specific Slice-as-a-Service, Virtual Network Function and integrated digital Platform-as-a-Service (dPaaS) marketplaces have arisen as all-in-one solutions.
In contrast, the aPaaS and iPaaS approaches are API-centric and focus on support and integration of well-defined standard APIs. The difference and overlap between the two approaches are shown in Fig.~\ref{fig:platforms}. It is important to note that the iPaaS approach does not include the runtime environments within the scope of the service offering directly. Hence, it is feasible to deploy services regardless of the specific runtime environment in which they are executed. This enables flexibility and opportunities for optimizing where and how the services are executed independently of the iPaaS environments. So, it is proposed that marketplace integration should support the iPaaS features, with the optional extensions towards aPaaS and dynamic microservice (i.e. edge/cloud) runtimes, using deploy agents to provide both functional and non-functional related service monetization. Deployment agents from cloud vendors or third-party integrators facilitate multi-stakeholder collaboration among DTs, applications, services, and RIC vendors (open-source/proprietary).

The essence of the iPaaS lies in API publishing and subscription, with fine-grained access control and monitoring. This enables individual applications to request access to APIs, with adjustable throttling aligning with monetization guidelines that oversee the strategies within the marketplace. 
This process is facilitated through the utilization of JSON web tokens (JWTs), incorporating crucial details such as the subject, audience, issuer, and client claim or consumer key (azp/aud). Moreover, it encompasses scope and expiry information to ensure secure and controlled access for each request. API gateways validate tokens and retrieve linked subscription details to authorize and manage access. This facilitates reconciliation, assessment of performance levels, and accurate charging processes. 
These gateways are built on the open-source Choreo connect implementation, utilizing Envoy proxies (available on GitHub), and they are deployed within each Kubernetes cluster, serving as both ingress controllers and gateways for non-Kubernetes deployments.
In our test deployments, Kubernetes collaborates with a federated Consul service mesh to establish secure, direct east-west interactions among services. The standardized APIs exposed by test services and used for testing purposes are (1) A1-PPolicyManagementService (A1-P non-realtime RIC); (2) A1PESimulator (A1-P near-realtime RIC); (3) EnrichmentDataservice (A1-EI); (4) GenericDmaapandKafkaInformationProducer (DMaaP).
    
\begin{figure}[t]   
        \centering   
        \includegraphics[width=1\columnwidth]{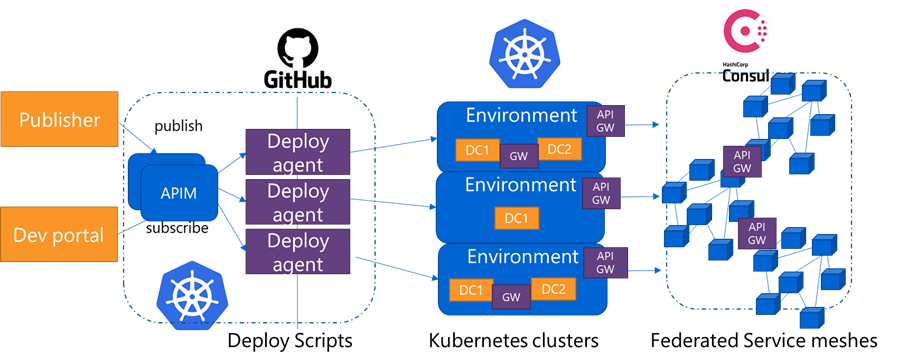}   \vspace{-5.00mm}
            \caption{Deployment process from API Manager.} \label{fig:deploy}
            \vspace{-5.00mm}
\end{figure}
\subsection{Marketplace Integration}
In order to select and deploy DT/applications/services, and monetise their use, a marketplace integration has been developed using the WSO2 API management-based iPaaS framework. We show the marketplace integration  features using the DT as an example. The DT are integrated and deployed from the marketplace publisher portal using a range of deployment options. This facilitates the automated deployment of DT across different Open RAN environments, using the corresponding deploy agents for each domain environment that are provided by the vendors. For testing purposes, a marketplace platform hosted on AWS at \url{https://beacon.umbrellaiot.com:9443} has been used. This approach enables the provision, selection, and deployment of diverse DT models, aligning with varying monetization policies, such as pay-per-use, performance-based criteria, service level agreements, as well as flat subscription fees. 
Employing an API-centric marketplace approach based on standardized Open RAN APIs facilitates these different options in a plug-and-play manner. For instance, A1-EI enrichment APIs and DMaaP-based APIs enable the acquisition of data for model training. Also, the forthcoming standardization of the A1-ML API will enable efficient model management.
Fig.~\ref{fig:deploy} illustrates the API Manager (APIM) marketplace that permits deployment agent plugins to access configurations from GitHub in order to deploy the DT to the different Open RAN Kubernetes environments when they are subscribed to.  In addition, a federated service mesh is utilised to facilitate the DT integration across different cloud and edge data centre environments in a secure and generic manner, supporting the interactions between the different training and inference parts of the DT.   

\begin{figure}[t]   
        \centering   
        \includegraphics[width=0.9\columnwidth]{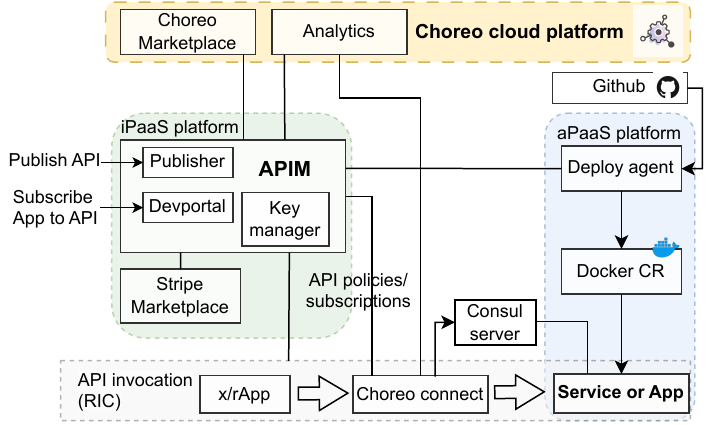}   
            \caption{Overall integration architecture with API invocation.} \label{fig:architecture}
            \vspace{-5.00mm}
\end{figure}
\subsection{Example DT Integration}
The purpose of the example 5G DT is to model the RAN elements, using a simulation environment, in order to permit a comparison of performance with the real Open RAN system. This can help with the automated selection of optimal scheduling schemes and also the detection and identification of the cause of performance differences. 
There are two deployment options that are considered to achieve the goals. The first is that the DT deployment package is self-contained and performs an automated scheduling selection policy as an output. The second is when the DT package outputs the performance and another service or rApp makes a policy selection based on the DT's output. 
In the case that the DT performs policy selections, the DT requires the actual performance from the RAN (through A1-EI APIs) and the output is a policy update towards the corresponding OSC, ONF or proprietary RIC instances. However, in the case that this operation is performed by a separate rApp then the output is per UE performance parameters that are exposed through the A1-EI API, which can be supported in addition to the A1-P policy API.
\subsection{Analytics Monitoring and Reconciliation}
The monitoring and reconciliation or auditing and billing functions are essential parts of the marketplace. These are provided as three possible options. Firstly, using the Stripe marketplace integration plugins and Choreo analytics server that are supported in the WSO2 APIM, and secondly using the centralised Choreo marketplace that is supported by the Choreo connect microgateways. Finally, there is also the option to utilise a fully decentralised auditing and reconciliation framework based on a Hyperledger blockchain digital ledger technology fabric implementation. 

\section{Remarks}
\vspace{-1.00mm}
This work demonstrated integration and application Platform-as-a-Service based marketplace strategy for effective integration and monetization process among stakeholders within the Open RAN ecosystem.

\bibliographystyle{IEEEtran} %

\bibliography{IEEEabrv,references} 

\end{document}